\documentclass[twocolumn,showpacs,preprintnumbers,prb,aps,superscriptaddress]{revtex4-2}
\usepackage{amsmath}
\usepackage{amssymb}
\usepackage[retain-unity-mantissa=false]{siunitx}
\usepackage{graphicx}
\usepackage{xcolor}
\usepackage[caption=false]{subfig}
\usepackage{placeins}
\usepackage[colorlinks=true, linkcolor=blue, citecolor=blue, filecolor=blue, urlcolor=blue,pdfproducer={.},pdfcreator={.}]{hyperref}
\usepackage{cleveref}
\usepackage{blindtext}
\usepackage{comment}
\usepackage{tikz}
\usepackage{verbatim}
\usepackage{ulem}
\usepackage{physics}
\usepackage{orcidlink}

\newcommand{\niceint}[2]{\int #1 \mathrm{d}#2 } %nicer looking integral
\renewcommand{\eqref}[1]{equation~(\ref{#1})}
\newcommand{\eV}{\electronvolt}
\newcommand{\fs}{\femto\second}
\newcommand{\nm}{\nano\meter}

\begin{document}

	\title{Competing thermalization pathways of photoexcited hot electrons}
	\author{Christopher \surname{Seibel} \orcidlink{0000-0003-1513-1364}}
	\email{cseibel@rptu.de}
    \author{Tobias Held \orcidlink{0009-0009-8925-1810}}
    \author{Markus Uehlein \orcidlink{0000-0002-3193-3749}}
	\author{Baerbel \surname{Rethfeld} \orcidlink{0009-0008-9921-4127}}
	\affiliation{Department of Physics and Research Center OPTIMAS, RPTU University Kaiserslautern-Landau, 67663 Kaiserslautern, Germany}
	
	\begin{abstract}
    Photoexcited hot carriers in solids can drive processes, such as photocatalytic reactions on the surface, beyond those available in thermal equilibrium.
    Hot-electron-mediated reaction pathways are limited by the thermalization of the nonequilibrium electron distribution through microscopic scattering events.
    Commonly, thermalization is exclusively attributed to electron-electron scattering, whereas electron-phonon scattering is considered relevant mainly for the energy equilibration with the lattice.  
    With a kinetic model based on full Boltzmann collision integrals, we demonstrate that each scattering mechanism alone can thermalize the electron distribution, albeit along different trajectories in phase space.
    We find an opposite dependence on the excitation strength of the respective thermalization times and show that both processes can become comparable for weak excitations, corresponding to a sample temperature increase of a few Kelvin. 
    Our results unravel the contributions of electron-electron and electron-phonon scattering to the thermalization across the full range of experimental excitation strengths up to the melting regime, thus facilitating the prediction of thermalization times for hot-carrier-based applications. 
	\end{abstract}

	\date{\today}

	\maketitle

    \section{Introduction}

    Light-matter interaction is pivotal for many applications ranging from material processing and medical surgery to solar energy harvesting with photocatalysis or solar cells~\cite{Wu2015, Sousa2016,Ivanov2015,Chichkov1996,Lubatschowski2000,Tang2020}. 
    Irradiating solids with ultrashort light pulses initiates a series of complex processes that provide access to fundamental mechanisms on their intrinsic timescales~\cite{Heide2024,Caruso2026,Khurgin2024,Bennemann2004}. 
    The impinging optical photons drive the initially Fermi-distributed electrons out of equilibrium, while the lattice remains cold. 
    Subsequently, the electrons thermalize towards a Fermi distribution of elevated temperature and transfer energy to the cold lattice until the electronic and phononic temperatures are equal.
    Depending on the considered sample geometry, ballistic and diffusive transport processes contribute to the energy distribution in the system~\cite{Hohlfeld1997,Rethfeld2017,Hopkins2009b}.
    These nonequilibrium dynamics directly determine the transient optical properties of solids and thus govern their ability to, in turn, manipulate light on ultrashort timescales~\cite{Lei2024,Ndione2022,Jiang2018}.

    The thermalization of hot non-thermal electrons towards a Fermi distribution with elevated temperature has been intensively studied over the past decades because it is among the first processes following ultrafast optical excitation.
    Its timescale influences subsequent processes and limits, for instance, the transfer of hot carriers to adjacent semiconductors, 2D materials such as transition metal dichalcogenides (TMDs), or molecules relevant for applications~\cite{Wu2015, Pincelli2023, Gao2025}.
    Advances in laser and measurement technology made it possible to follow the temporal evolution of a laser-excited nonequilibrium electron distribution in gold with time-resolved photoelectron spectroscopy in the early 1990s~\cite{Fann1992a,Fann1992b}.
    This pioneering experiment triggered extensive investigation of thermalization dynamics across a wide range of materials and excitation conditions~\cite{Bauer2015,Petek1997,Rethfeld2002,Bovensiepen2010,Dubi2019,Jiang2018}.
    For typical metals, thermalization times ranging from a few femtoseconds to several hundred femtoseconds have been reported, depending on excitation conditions~\cite{Medvedev2011,Inogamov2010,Petrov2016,Sun1994,Guo2001}.

    Microscopically, various scattering processes govern the dynamics of electrons and phonons following laser excitation. 
    It is often assumed that electron-electron scattering leads to the thermalization of electrons on femtosecond timescales, whereas electron-phonon collisions mainly drive the energy transfer to the lattice and thus the equilibration of electron and lattice temperatures on picosecond timescales~\cite{Schirato2023,Bauer2015,Mathias2012,Rethfeld2017}. 
    Due to these different timescales, the two processes are often considered temporally ordered and mostly independent, and the influence of phonons on the thermalization of electrons is thus frequently neglected. 
    However, the difference in their macroscopic timescales mainly originates from the different energy transfer efficiency of the individual collisions~\cite{Rethfeld2002}, while the scattering frequencies of both processes are comparable~\cite{Bejan1997,Lugovskoy1999} or even larger for electron-phonon scattering~\cite{Bernardi2015}.
    For semiconductors, it has recently been shown that the dynamics of hot photocarriers on early timescales cannot be fully captured without considering both electron-electron and electron-phonon scattering~\cite{Mocatti2025arxiv}, even though in this case the former is often neglected~\cite{Tong2021,Caruso2021}.
    Also for metals, there are indications that the thermalization times arising from electron-electron and electron-phonon scattering might overlap for weak excitations~\cite{Obergfell2020,Kratzer2022,Kabanov2008,Baranov2014}. 
    Therefore, a systematic investigation of thermalization times in dependence on excitation strength and the role of phonons in the thermalization is required. 

    Theoretical modeling is a valuable tool for studying electron-electron and electron-phonon scattering independently, since experimentally the roles of these processes are hard to disentangle. 
    Models based on the Boltzmann equation have proven particularly suitable to describe the dynamics of laser-excited electron distributions resulting from microscopic scattering processes~\cite{Rethfeld2002,Kabanov2008,Mueller2013PRB,Nenno2016, Dubi2019, Seibel2023,Caruso2022,Weber2025}.

    Here, we use a kinetic model based on full Boltzmann collision integrals to study the influence of electron-electron and electron-phonon scattering on the thermalization of laser-excited metals. 
    We evaluate the temporal evolution of the electronic distribution of an aluminum-like system, considering both processes separately as well as simultaneously. 
    The analysis reveals different thermalization pathways of the individual processes and an opposite dependence of their respective timescale on the excitation strength. 
    In particular, we explicitly show that the thermalization times associated with electron-electron and electron-phonon scattering become comparable for weak excitations and demonstrate that the microscopic processes are strongly intertwined on ultrashort timescales.

    \section{Results and Discussion}
    
    We numerically calculate the thermalization behavior of the electrons after ultrashort laser excitation based on Boltzmann collision integrals for the microscopic scattering processes, see the Methods section for details. 
    To isolate the role of the scattering channels in the relaxation dynamics, we choose a free electron gas-like density of states (DOS) $D(E) \propto \sqrt{E}$. 
    The band mass and Fermi energy are chosen to resemble aluminum, whose DOS is close to that of a free electron gas~\cite{Lin2008,Uehlein2025}.
    For the phonons, we assume a triply degenerate Debye dispersion with a speed of sound of \SI{6420}{\meter\per\second}~\cite{CRC2005}.

    After optical laser excitation, the electrons are in a state far from equilibrium and their energy distribution does not follow Fermi-Dirac statistics. 
    We define the corresponding equilibrium distribution $f_{\text{eq}}$ as the Fermi distribution with the same energy and particle content as the respective nonequilibrium distribution $f_{\text{neq}}$~\cite{Mueller2013PRB,Uehlein2025}. 
    To quantify the deviation from equilibrium, we use the mean absolute deviation (MAD) between these distributions~\cite{Roden2026arxiv}
    \begin{equation}
        \Delta(t) = \niceint{\abs{f_\text{neq}(E,t) - f_\text{eq}(E, T(t), \mu(t))}}{E} . 
        \label{eq:mad}
    \end{equation}
    Note that if energy or particles are exchanged with other systems during thermalization, as is the case for electron-phonon coupling, the corresponding equilibrium distribution becomes time-dependent through the evolving temperature $T$ and chemical potential $\mu$.

\begin{figure}
        \centering
        \includegraphics[width=\linewidth]{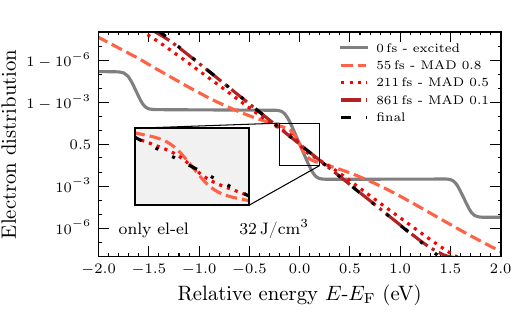}
        \includegraphics[width=\linewidth]{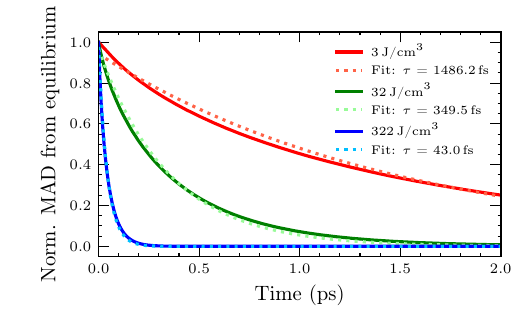}
        \caption{Thermalization purely mediated by electron-electron scattering. a) Electron distributions at different stages of the thermalization, defined by the mean absolute deviation (MAD) from equilibrium, for an excitation strength of \SI{32}{\joule\per\centi\meter\cubed}. The gray distribution indicates the initial state after excitation that will eventually relax to the final thermal distribution (black dash-dotted). Note the symmetric logarithmic scale. b) Temporal evolution of the mean absolute deviation from the equilibrium distribution for three different excitation strengths. The thermalization times are determined by means of an exponential fit.}
        \label{fig:dist_el}
    \end{figure}

\begin{figure}
        \centering
        \includegraphics[width=\linewidth]{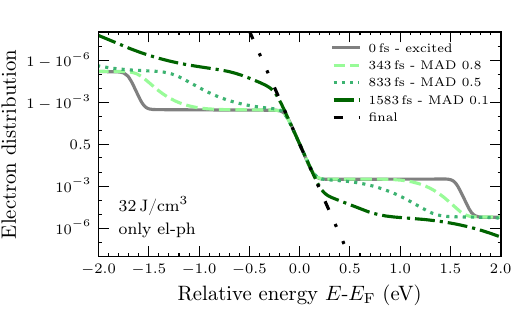}
        \includegraphics[width=\linewidth]{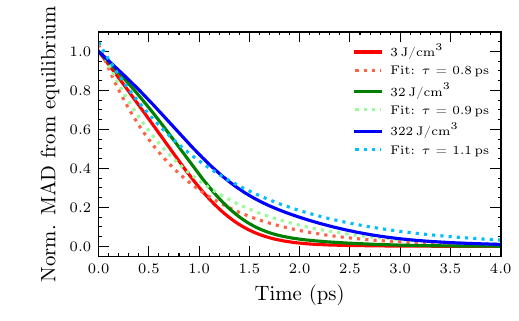}
        \caption{Same as \cref{fig:dist_el}, with pure electron-phonon scattering mediating thermalization.}
        \label{fig:dist_ph}
    \end{figure}

\begin{figure}
        \centering
        \includegraphics[width=\linewidth]{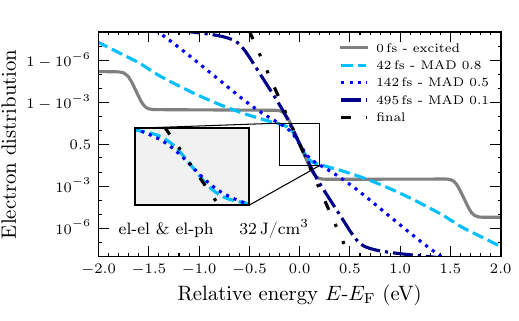}
        \includegraphics[width=\linewidth]{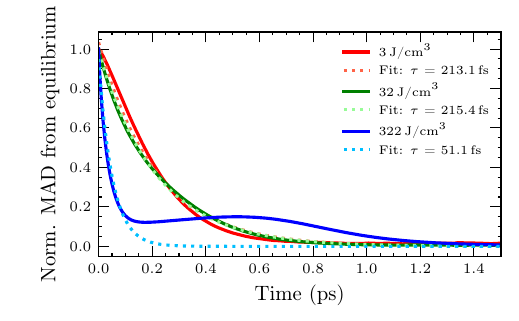}
        \caption{Same as \cref{fig:dist_el}, but with both electron-electron and electron-phonon scattering included for the thermalization.}
        \label{fig:dist_elph}
    \end{figure}

    To disentangle the influence of the microscopic processes on thermalization, we consider three scenarios and perform simulations including i) pure electron-electron scattering, ii) pure electron-phonon scattering, and iii) both processes acting simultaneously.
    \Cref{fig:dist_el,fig:dist_ph,fig:dist_elph} show the corresponding electron distributions and thermalization dynamics. 
    
    In all simulations, the system is excited by a laser pulse at a wavelength of \SI{800}{\nm} corresponding to a photon energy of $\hbar\omega=\SI{1.5}{\eV}$. 
    Parts a) of \cref{fig:dist_el,fig:dist_ph,fig:dist_elph}
    show the initial nonequilibrium distribution (solid gray line) resulting from an excitation with an absorbed energy density of \SI{32}{\joule\per\centi\meter\cubed}.
    It exhibits a characteristic step-like structure with the width of the steps corresponding to the photon energy. 
    This excitation corresponds to a peak electron temperature of \SI{873}{\kelvin} and increases the final sample temperature by about \SI{13}{\kelvin}.
    To separate thermalization from excitation, the thermalization mechanisms begin to act only after the laser pulse has ended, defining time zero. 
    
    For each case, we also show the final Fermi distribution after thermalization (black dash-dotted line). In the symmetric logarithmic representation, this distribution appears as a straight line whose slope is determined by the temperature~\cite{Rethfeld2002}.

    To compare the relaxation pathways despite their widely different timescales, we evaluate the distributions at fixed values of the normalized MAD $\Delta(t)/\Delta(0)$, namely 0.8, 0.5, and 0.1.
    At a given MAD, the electron systems exhibit the same degree of nonequilibrium, thus the different processes have led to the same degree of thermalization, which facilitates a direct comparison of the underlying microscopic pathways in \cref{fig:dist_el,fig:dist_ph,fig:dist_elph}.

    \Cref{fig:dist_el} shows the thermalization dynamics when only electron-electron scattering is included. 
    Already at an MAD of 0.8, reached here after \SI{55}{\fs}, the step-like structure of the excited distribution has been largely washed out.
    However, the population above the Fermi level is still substantially larger than in the final equilibrium state. 
    Close to the Fermi level, a residual step edge is still visible. 
    This slower thermalization around the Fermi level is qualitatively consistent with Fermi-liquid theory.
    As thermalization progresses, the distribution also straightens around the Fermi edge and electrons from above the Fermi level fill holes below it. 
    Already at an MAD of 0.5, the distribution appears to be already almost thermal, i.e. similar to the final one, over a broad energy range around the Fermi level.
    Close to equilibrium, at an MAD of 0.1, the distribution is essentially indistinguishable from the final Fermi distribution. 

    The observed thermalization behavior originates from electron-electron scattering being a long-range interaction in energy space. 
    Thus, electrons equilibrate globally and the entire system jointly moves towards the final Fermi distribution. 
    
    In addition to the distribution snapshots in \cref{fig:dist_el}\,a),  \cref{fig:dist_el}\,b) shows the temporal evolution of the MAD for three different excitation strengths, resulting in electron temperatures of \SI{396}{\kelvin}, \SI{873}{\kelvin}, and \SI{2610}{\kelvin}, respectively.
    Thermalization proceeds faster for stronger excitation, in agreement with previous findings~\cite{Mueller2013PRB, Fann1992b,Lisowski2004}.
    Exponential fits yield time constants between \SI{1.5}{\ps} and \SI{43}{\fs}.
    The close agreement between fit and simulation indicates that the integrated nonequilibrium decays almost exponentially when thermalization is mediated solely by electron-electron scattering.
    This behavior, however, characterizes only the global measure of nonequilibrium, whereas the energy-resolved dynamics can still be strongly non-exponential and involve multiple timescales~\cite{Seibel2023,Roden2026arxiv}.

    The thermalization behavior changes qualitatively when only electron-phonon scattering is taken into account, as revealed in \cref{fig:dist_ph}\,a).
    The excitation conditions are identical to those in \cref{fig:dist_el}\,a), so the initial excited distribution (solid gray line) is the same.
    However, the final distribution has a lower temperature (steeper slope) than in \cref{fig:dist_el}\,a) because most of the energy is transferred to the phonons during thermalization. 
    The final temperature is only slightly larger than the temperature before excitation, which nearly preserves the initial slope before excitation around the Fermi edge~\cite{Rethfeld2002}.
    
    In this case of pure electron-phonon scattering, an MAD of 0.8 is reached after \SI{343}{\fs}, and thus much later than in the case of electron-electron mediated thermalization.
    At this time, the distribution near the Fermi edge has not changed compared to the excited one.
    Likewise, the plateaus of the first step above and below the Fermi edge remain essentially intact, in contrast to the electron-electron scattering discussed above.
    The main changes occur near the step edges located one photon energy, \SI{1.5}{\eV}, above and below the Fermi level, where electron-phonon scattering has slightly smoothed the edges.

    As the thermalization proceeds, this smoothing continues and the decay of the steps moves progressively toward the Fermi edge, as seen at an MAD of 0.5.
    Close to equilibrium, at an MAD of 0.1 after \SI{1.5}{\ps}, visible deviations from the final Fermi distribution remain.
    This differs markedly from the electron-electron scattering case in \cref{fig:dist_el}, where the distribution in the displayed range is already almost indistinguishable from equilibrium at the same MAD.
    Note that also in the case of electron-phonon scattering in \cref{fig:dist_ph} the difference to equilibrium is small, however, the low density regions appear exaggerated in the logarithmic representation. 
    
    The evolution of the distribution during the thermalization with pure electron-phonon coupling differs qualitatively from that driven by electron-electron scattering. 
    As phonon energies are on the \SI{}{\milli\eV} scale compared to the \SI{}{\eV} scale of electrons, electron-phonon collisions redistribute energy less efficient than electron-electron collisions.
    Electron-phonon scattering is thus short-ranged in the energy space of the electrons and thermalization proceeds locally rather than globally.

    The time evolution of the MAD for this case is shown in \cref{fig:dist_ph}~b).
    The nonequilibrium decays on a timescale of a few ps, demonstrating that phonons alone can thermalize electrons in metals even in the absence of electron-electron scattering.
    In contrast to the behavior found above, here the decay is more complex than exponential.
    After a slower onset, the thermalization accelerates for all three considered excitation strengths.
    For comparison with the electron-electron case, we fit the dynamics with a single exponential.
    The resulting time constants lie on the picosecond scale and, opposite to the electron-electron case, increase with excitation strength. 
    
    \Cref{fig:dist_elph}\,a) shows the evolution of the same initial excited distribution as above, but for the case when electron-electron and electron-phonon scattering act simultaneously.
    Here, an MAD of 0.8 is reached after \SI{42}{\fs} and thus faster than for both individual scattering processes, respectively. 
    At this time, the step-like structure of the excited distribution has disappeared, except for a slight edge at the Fermi energy. 
    In this respect, the dynamics resemble the case of pure electron-electron scattering.
    Farther away from the Fermi edge, however, the distribution is less straight than in \cref{fig:dist_el}\,a).
    As before, the excited and final distributions have a similar slope near the Fermi edge because the weak excitation increases the temperature only slightly once energy transfer to the phonons is taken into account.

    Later during the thermalization, at an MAD of 0.5, the distribution far from the Fermi edge straightens further and tilts towards the final distribution because of electron redistribution. 
    At the same time, the energy range near the Fermi edge where the distribution matches the final distributions shrinks, as can be seen from the difference between the light blue dashed and the blue dotted line in the region $E_\text{F} \pm \SI{0.2}{\eV}$ in the inset. 
    This implies that nonequilibrium electrons far from the Fermi edge drive electrons near the Fermi level
    towards a transient equilibrium of different temperature via electron-electron collisions.
    This resembles the conditions underlying a two-temperature description. 

    Even close to the thermal state, at an MAD of 0.1, the distribution still has a slightly smaller slope than the final Fermi distribution.
    This differs from the pure electron-electron case in \cref{fig:dist_el}\,a). 
    Moreover, in the low-density range below $f = \num{1e-6}$ or above $1 - \num{1e-6}$ plateaus are visible which are reminiscent of the behavior found for pure electron-phonon scattering. 
    Over the full energy range displayed, the agreement with the final state is better than for pure electron-phonon scattering in \cref{fig:dist_ph}\,a).

    A direct comparison of \cref{fig:dist_el} and \cref{fig:dist_elph} shows that electron-phonon scattering plays an important role in the thermalization, even though this process is often assumed to be governed by electron-electron scattering alone.
    While both types of scattering are present, they contribute to the thermalization through distinct pathways, reflecting their different ranges in energy space: electron-electron scattering promotes global equilibration, whereas electron-phonon scattering favors local relaxation.
   
    \Cref{fig:dist_elph}\,b) shows the decay of the nonequilibrium in the system. 
    Unlike for the individual scattering channels, the thermalization times no longer exhibit a simple monotonic dependence on the absorbed energy.
    For the two lowest excitation strengths, the thermalization time is almost the same with a slight increase, while for the strongest excitation it decreases significantly.
    In particular for the largest absorbed energy, the MAD shows a strongly non-exponential decay.
    A rapid initial decrease is followed by an almost plateau-like regime with a much slower decay. 
    This behavior is at least partly caused by the continuous transfer of energy from electrons to phonons, which changes the equilibrium distribution entering the definition of the MAD, see \cref{eq:mad}.
    As a consequence, a single exponential fit captures mainly the initial fast stage and underestimates the total relaxation time.
    The actual thermalization is therefore slower than suggested by the fitted timescale, in particular for strong excitation.
    Nevertheless, we retain the fit for consistency with the previous cases.

    The coupled dynamics resulting from electron-electron and electron-phonon scattering thus give rise to a more intricate thermalization behavior than either process alone.
    The slowing down of the thermalization after a fast initial decay was already previously observed, especially in regions with drastic changes of the DOS and explained by phonons driving electrons out of equilibrium~\cite{Weber2019}. 
    The delayed thermalization indicates that electron-electron and electron-phonon scattering compete with each other, which will be discussed in more detail below.

    \begin{figure}
        \centering
        \includegraphics[width=\linewidth]{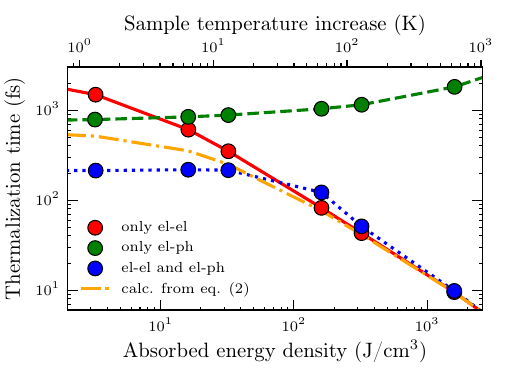}
        \caption{Comparison of the fitted thermalization times for various excitation strengths. At very weak excitations, the thermalization times due to electron-electron and electron-phonon scattering become comparable.}
        \label{fig:relaxation_time_comp}
    \end{figure}

    In \cref{fig:relaxation_time_comp}, we compare the fitted relaxation times extracted from the dynamics of the different cases discussed in \cref{fig:dist_el,fig:dist_ph,fig:dist_elph}\,b) over a broad range of excitation strengths.
    For reference, the top axis indicates the increase of the sample temperature with respect to the initial temperature of \SI{300}{\kelvin} reached after electrons and phonons have equilibrated.
    Note that the temperature reached by the electrons after the thermalization in case of only electron-electron scattering is much higher because the energy transfer to the lattice has yet to take place.
    The observed range of few fs to several hundred fs for the thermalization depending on the excitation strength agrees well with previous theoretical~\cite{Inogamov2010,Medvedev2011,Petrov2016,Rethfeld2002,Gusev1998} and experimental studies~\cite{Bauer2015,Sun1994,Fann1992b,Guo2001} for free electron-like metals.
    
    As already observed above, the thermalization times for the case when only the electron-electron interaction is considered decrease drastically with increasing excitation strength, whereas the thermalization is much slower when only electron-phonon interaction is considered. 
    In general, electron-phonon-driven thermalization is slower and much less fluence dependent than the electron-electron-driven one. 
    For the weakest considered excitations, the electron-phonon-driven thermalization is comparably fast as or even faster than the thermalization mediated only by electron-electron scattering. 
    The existence of such a regime is indicated by the opposite monotonic trends in \cref{fig:relaxation_time_comp} and has been suggested previously~\cite{Obergfell2020,Kratzer2022,Kabanov2008,Baranov2014}.

    Electron-electron and electron-phonon scattering are often treated as independent processes.
    Under this assumption, the combined thermalization time could be determined from the individual electron-electron thermalization time $\tau_{ee}$ and electron-phonon-mediated thermalization time $\tau_{ep}$ by 
    \begin{equation}
        \frac{1}{\tau} = \frac{1}{\tau_{ee}} + \frac{1}{\tau_{ep}}.
        \label{eq:indep_thermalization_time}
    \end{equation}
    The resulting thermalization time $\tau$ is shown in \cref{fig:relaxation_time_comp} as yellow dash-dotted line.
    However, the actual thermalization times obtained when both scattering channels are included differ significantly from this prediction.
    For small absorbed energies, the combined dynamics are much faster than expected for independent processes.
    In this regime, the pathways identified in \cref{fig:dist_el,fig:dist_ph} act cooperatively and accelerate thermalization.
    In contrast, at larger absorbed energy densities the combined dynamics are slower than the superposition of the individual processes. 
    Here, the processes compete, and the presence of the phonons retards the thermalization. 
    This retardation may in fact be even stronger than suggested by the fitted times, since the exponential fit does not capture the full dynamics for strong excitation, as discussed for \cref{fig:dist_elph}\,b).
    As long as energy is continuously transferred to the phonons, the electrons do not reach a fully stationary thermal state. 
    Electron-electron and electron-phonon scattering therefore cannot, in general, be treated as independent, but are strongly entangled.
    The degree of their (in-)dependence critically depends on the excitation strength. 
    
    For the extreme case of very strong excitation, approaching or exceeding the melting threshold relevant for ablation and nanostructuring applications~\cite{Mo2018,Sun2025,Bonse2023}, the phonon contribution to the thermalization dynamics can be neglected and the assumption of independent electron-electron and electron-phonon scattering is justified.
    
    For the other extreme case of weak excitations, below about \SI{50}{\joule\per\centi\meter\cubed} and corresponding to temperature increases of about \SI{20}{\kelvin} for our system, phonons must be included in the description of electron thermalization.
    Such conditions are common in time-resolved photoemission experiments~\cite{Fann1992b,Kuehne2022,Knorren2000}, for slight lattice deformations~\cite{Brand2025,Li2018}, or plasmonic nanostructures with applications in photocatalysis~\cite{Gao2025,Boerigter2016,Khurgin2024}.
    Their interpretation therefore requires explicit consideration of phonon-assisted electron thermalization. 
    
    At intermediate excitation strengths, such as those used in studies of ultrafast magnetization dynamics~\cite{Hofherr2020,Tengdin2018} and reversible structural dynamics~\cite{Mo2018b,Waldecker2016,Pudell2018}, most of the thermalization occurs via electron-electron scattering on short timescales. 
    However, electron-phonon coupling can maintain a partial nonequilibrium in the electronic system over longer times~\cite{Weber2019}. 

    The excitation thresholds separating these regimes strongly depend on the material parameters. 
    The aluminum-like system considered here combines strong electron-phonon coupling with a low electronic heat capacity, which strongly contributes to the comparable thermalization times of electron-electron and electron-phonon scattering at weak excitations. 
    In materials such as gold, where electron-phonon coupling is much weaker, phonon-induced thermalization is expected to be substantially slower. 
    In that case, the approximation of independent scattering mechanisms should remain valid over a wider range of excitation strengths.
    The degree of independence between electron-electron and electron-phonon scattering must therefore be assessed separately for each material.

    \section{Conclusion}

    We have calculated the evolution of the electron energy distribution during thermalization in an aluminum-like model system for different microscopic processes based on full Boltzmann collision integrals. 
    
    We considered three different scenarios, namely pure electron-electron scattering, pure electron-phonon scattering, and the combined processes driving thermalization. 
    By comparing the temporal evolution of the electronic distribution, we have shown that electron-electron and electron-phonon scattering act differently during the thermalization stage.
    The energetically long-ranged electron-electron interaction leads to a global thermalization, whereas the short-ranged electron-phonon interaction thermalizes the electrons locally in energy space. However, both play an important role for the thermalization.
    In particular, phonons, which are usually disregarded in the thermalization of metals, can significantly contribute to thermalization and even theoretically fully thermalize the electron system in absence of electron-electron scattering. 
    We have shown that, depending on the excitation strength, these two processes can either cooperate or compete. For weak excitations, their interplay accelerates the thermalization, whereas it hinders it for stronger excitations.
    In the former regime, the timescales of electron-phonon and electron-electron scattering become comparable for the considered system.
    Thus, especially in this regime but also for stronger excitations, electron-electron and electron-phonon scattering are strongly intertwined and cannot be treated as independent processes.

    In conclusion, our results disentangle the microscopic pathways governing electron thermalization in laser-excited metals across a wide range of excitation strengths.
    They thereby enable more reliable estimates of carrier lifetimes relevant for hot-electron-based applications.
    
    \section{Methods}

    We describe the nonequilibrium dynamics of electrons and phonons after laser excitation with an established method based on the quantum Boltzmann equation~\cite{Rethfeld2002,Mueller2013PRB,Seibel2023,Roden2026arxiv}.
    Within this framework, the temporal evolution of the electron and phonon energy distributions, $f(E,t)$ and $g(E,t)$, is given by
    \begin{align}
        \dv{f(E,t)}{t} &= \left.\pdv{f}{t}\right|_{\text{el-el}}
        + \left.\pdv{f}{t}\right|_{\text{el-ph}}
        + \left.\pdv{f}{t} \right|_{\text{laser}} \\
        \dv{g(E,t)}{t} &= \left.\pdv{g}{t}\right|_{\text{ph-el}},
    \end{align}
    where the subscripts denote the considered electron-electron scattering, electron-phonon scattering, and laser excitation for the electrons, and phonon-electron scattering for the phonons. 
    All processes are described by full Boltzmann collision integrals of similar structure, involving a transition matrix element that determines the scattering probability, and a collision functional which ensures Pauli's principle. 

    The electron-electron collision integral is evaluated in the random-\textbf{k} approximation, and a plane wave matrix element of a screened Coulomb potential is considered for the transition probability~\cite{Seibel2025CommPhys,Roden2026arxiv}. 
    Instead of a static Thomas-Fermi screening, we consider a dynamically evolving screening that depends directly on the electron distribution~\cite{DelFatti2000}.

    The collision integral for the laser excitation is based on inverse bremsstrahlung and accounts for Pauli blocking and multiphoton excitations~\cite{Rethfeld2002,Seibel2023}.

    For electron-phonon scattering, we include both phonon emission and absorption and rely on an analytical plane wave transition matrix element~\cite{Rethfeld2002,Mueller2013PRB,Held2025a}.
    The corresponding change of the phonon distribution is obtained in a separate collision integral for phonon-electron scattering~\cite{Rethfeld2002,Mueller2013PRB,Held2025c}.
    As the evolution of the phonon distribution is not of particular interest here, we determine the total internal energy change of the phonons from the collision integral $\left.\pdv{g}{t}\right|_{\text{ph-el}}$ and trace the corresponding temperature change, thus keeping the phonon system always in thermal equilibrium. 

\bigskip
    
    \section*{Acknowledgments}

    Financial support of the Deutsche Forschungsgemeinschaft (DFG, German Research Foundation) through the SFB/TRR-173-268565370 ‘Spin+X’ (Project No. A08) is gratefully acknowledged. 
    We appreciate the Allianz für Hochleistungsrechnen Rheinland-Pfalz for providing computing resources through project STREMON on the Elwetritsch high-performance computing cluster.
    
	\bibliography{main.bbl}

\end{document}